\documentclass[a4paper,12pt]{article}
\usepackage{bm}
\usepackage{amsmath}
\usepackage{amsfonts}
\usepackage{amssymb}
\usepackage{graphicx}
\usepackage{graphics}
\usepackage[dvips]{color}
\usepackage{anysize}

\marginsize{2.5cm}{2.0cm}{2.0cm}{2.0cm}
\newcommand{\be}{\begin{equation}}
\newcommand{\ee}{\end{equation}}
\newcommand{\ba}{\begin{eqnarray}}
\newcommand{\ea}{\end{eqnarray}}

\newcommand{\ubarf}{\overline{u}({\bf p}_N,s_N)}

\newcommand{\nk}{{\bf      k}}
\newcommand{\np}{{\bf      p}}

\newcommand{\nq}{{\bf      q}}

\title{Neutrino-Induced 1-$\pi$ Production}

\author{R. Gonz\'alez-Jim\'enez\footnote{raul.gonzalezjimenez@ugent.be}, T. Van Cuyck, N. Van Dessel, V. Pandey, N. Jachowicz}
\date{Department of Physics and Astronomy, Ghent University,\\ Proeftuinstraat 86, B-9000 Gent, Belgium.\\
\vspace{0.2cm}
November 2015}

\begin{document}
\maketitle

\begin{abstract}
Neutrino-induced pion production constitutes an important contribution to neutrino-nucleus scattering cross sections at intermediate energies. A deep understanding of this process is mandatory for a correct interpretation of neutrino-oscillation experiments.
We aim at contributing to the ongoing effort to understand the various experimental results obtained by different collaborations in a wide range of energies. In particular, in this work we analyze recent MiniBooNE and MINER$\nu$A charged-current neutrino 1-$\pi$ production data. 
We use a relativistic theoretical approach which accounts for resonant and non-resonant 1-$\pi$ production contributions. 
\end{abstract}

\vspace{1cm}


Most of the recent neutrino-nucleus scattering experiments work in the intermediate energy region (incident neutrino energy from some hundreds of MeV to a few GeV) where the quasielastic (QE) and the pion production channels are the dominant reaction mechanisms.
A good understanding of all possible channels involved in the reaction as well as accounting for nuclear effects is essential to obtain a reliable reconstruction of the neutrino energy and, consequently, to reduce systematic uncertainties in determining neutrino oscillation parameters. 
%
%
The recent measurements on neutrino-induced pion production reported by the MiniBooNE~\cite{MB11} and MINER$\nu$A~\cite{Minerva15} collaborations show that the current theoretical predictions are not able to reproduce the experimental cross sections with the desired accuracy. 
Additionaly, due to the increasing energies of the new generation of experiments (Hyper-Kamiokande and LBNF) a thorough understanding of the pion production process will be crucial in the near future.

\begin{figure}[tbh]
\centering
(a)\includegraphics[width=.2\textwidth,angle=0]{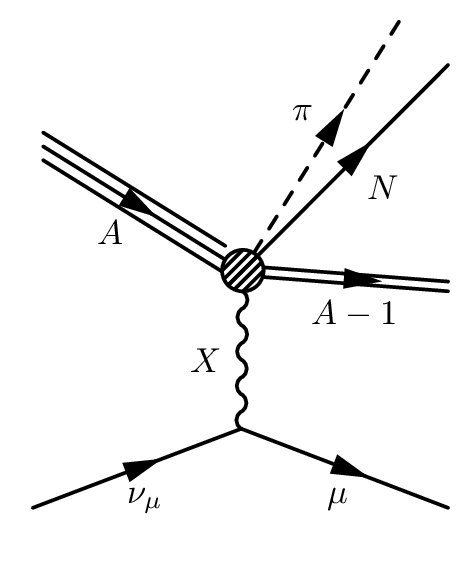}\hspace{0.5cm}
(b)\includegraphics[width=.55\textwidth,angle=0]{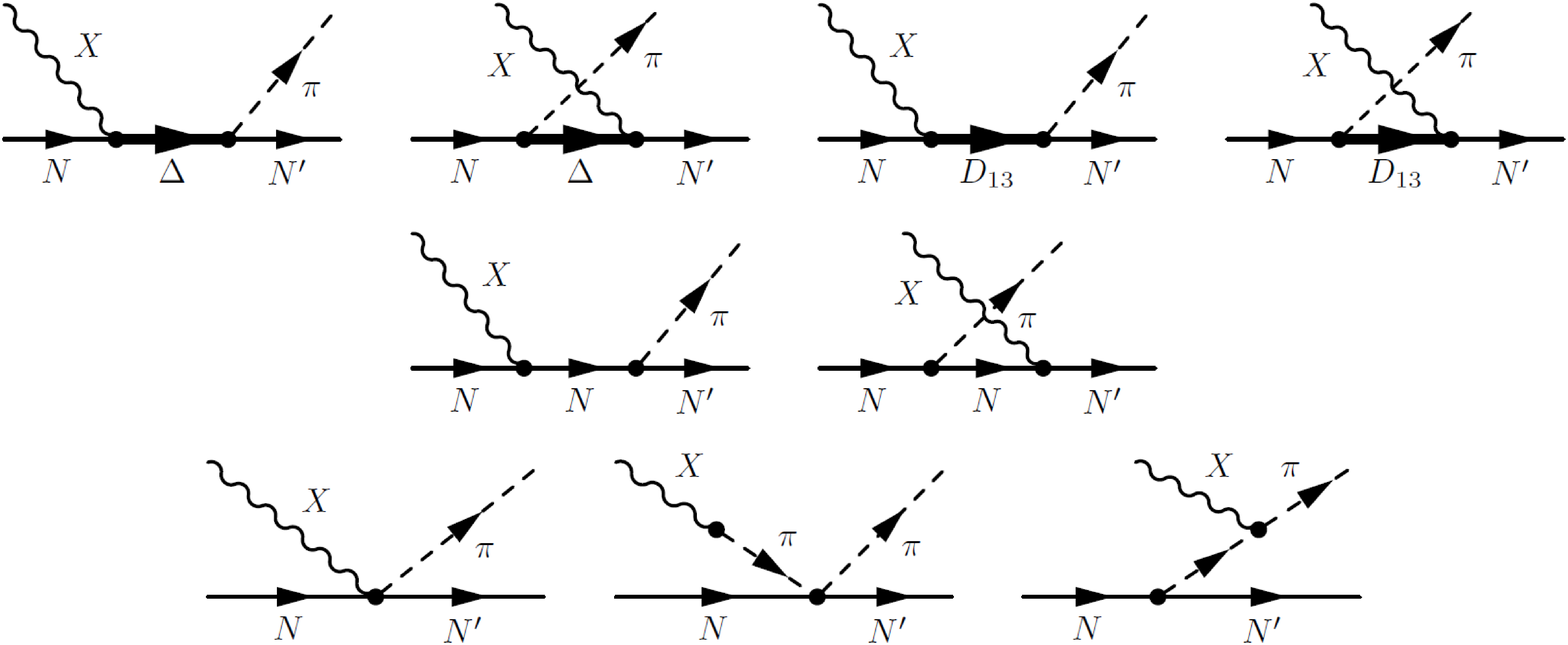}
\caption{(a) Charged-current neutrino-induced 1-$\pi$-production process. (b) Diagrams considered in this work for describing the 1-$\pi$ production~\cite{Hernandez07,Hernandez13}. From left to right and top to bottom: $\Delta$ pole ($\Delta$P), crossed delta pole (C$\Delta$P), $D_{13}$ pole (DP), crossed $D_{13}$ pole (CDP), nucleon pole (NP), crossed nucleon pole (CNP), contact term (CT), pion pole (PP) and pion-in-flight term (PF). In this work, $X$ represents the $W^+$ boson. }
\label{kinematic1}
\end{figure}
In this work, we focus on describing the process depicted in Fig.~\ref{kinematic1} (a). An incoming neutrino with 4-momentum $K_\nu^\mu(\varepsilon_\nu,\nk_\nu)$ interacts with a nucleus at rest $P_A^\mu(M_A,{\bf 0})$ by exchanging a single $W^+$ boson given by $Q^\mu(\omega,\nq)$.
The final state consists of a scattered muon $K_\mu^\mu(\varepsilon_\mu,\nk_\mu)$ in the leptonic sector and a nucleon $P_N^\mu(E_N,\np_N)$, a pion $P_\pi^\mu(E_\pi,\np_\pi)$, and the residual nucleus $P_{A-1}^\mu(E_{A-1},\np_{A-1})$ in the hadronic sector.
We use the impulse approximation (IA) to simplify the many-body problem, i.e., we assume that the $W^+$ boson couples to a single off-shell nucleon $P_i^\mu(E,\np)$ in the nucleus. For the description of the pion production we consider the nine diagrams shown in Fig.~\ref{kinematic1} (b): the excitation and decay of the $\Delta$~\cite{Praet09} and $D_{13}$ resonances~\cite{Hernandez13}, and the non-resonant contributions arising from the Lagrangian density of the non-linear sigma model as described in Ref.~\cite{Hernandez07}.


We build the cross section following Ref.~\cite{Praet09}. In the case of neutrino-nucleus scattering, 9 independent variables are needed to describe the scattering process. The cross section is proportional to the invariant matrix element resulting from the contraction of the leptonic ($\jmath^\mu_{\text{{\tiny lep}}}$) and hadronic currents ($J_{\text{{\tiny had}}}^\nu$):
\ba
\frac{d^8\sigma}{dE_\mu d\Omega_\mu dE_\pi d\Omega_\pi d\Omega_N}
  \propto  \overline{\sum} \left|{\cal M}_{fi} \right|^2 = 
  \overline{\sum}\left|\langle\jmath^\mu_{\text{{\tiny lep}}}\rangle\, S^W_{\mu\nu}\, \langle J_{\text{{\tiny had}}}^\nu \rangle \right|^2\,,\nonumber
\ea
where $S^W_{\mu\nu}$ represents the propagator of the $W^+$ boson. 
The leptonic current is described using Dirac plane waves for the leptons. 
In our model the nucleon bound-state wave function ${\cal U}_{\alpha,m}(\np_i)$ is obtained in the Hartree approximation to the Walecka model~\cite{Praet09} while the outgoing nucleon $\ubarf$ is a relativistic plane wave. 
Thus, within the IA, the hadronic current is given by
\ba
\langle J^\nu_{had}\rangle = \ubarf\ \hat{{\cal J}}^\nu\ {\cal U}_{\alpha,m}(\np_i)\,,\label{Jhad}
\ea
where $\hat{{\cal J}}^\nu$ represents the hadronic current operator which induces the transition between the initial 1-nucleon state and the final 1-nucleon 1-$\pi$ state. For this, we use the expressions given in Refs.~\cite{Hernandez07,Hernandez13} with the pion form factor in the NP, CNP, CT and PF terms 
as in Ref.~\cite{Sobczyk13}.
The process is described in a fully relativistic framework in which both kinematic and dynamic relativistic effects are taken into account, and in-medium corrections for the bound nucleon like Fermi motion, nuclear binding effects and Pauli blocking are naturally included. 
The outgoing pion and nucleon are described as plane waves.
Work on implementing final-state interactions (FSI) for the outgoing hadrons is in progress.
%
Finally, in the case of scattering off free nucleons, both the incoming and outgoing nucleons are relativistic plane waves and the cross section depends on 6 independent variables: $\frac{d^5\sigma}{dE_\mu d\Omega_\mu d\Omega_\pi}$.\\

It is well known that the $\Delta$ resonance is the dominant contribution to the 1-$\pi$ production process. 
The amplitude related to the dominant $\Delta$P term [Fig.~\ref{kinematic1} (b)] is given by 
\ba
\langle J^\mu_{\Delta}\rangle=\ubarf\,\, \Gamma_{\Delta\pi N}^{\alpha}\,\, S_{\Delta,\alpha\beta}\,\, \Gamma_{WN\Delta}^{\beta\mu}\,\,{\cal U}_{\alpha,m}(\np_i)\,.\nonumber
\ea
%
The $\Delta$ production vertex ($\Gamma_{WN\Delta}^{\beta\mu}$) is parametrized in terms of the $N\Delta$ transition form factors~\cite{Praet09,Hernandez07,Leitner09}.
For the $N\Delta$ transition vector form factors, we use the prescription presented in Ref.~\cite{Lalakulich06}, where the form factors were fitted to electroproduction helicity amplitudes. The $N\Delta$ transition axial form factors are tuned to fit BNL data~\cite{BNL86} for the $\nu_\mu + p \longrightarrow\mu^- + p + \pi^+$ channel when only the contribution from the $\Delta$-resonance is considered. Thus, in this work we use $C_5^A(0)$ = 1.2, $M_A=1.05$ GeV~\cite{Paschos04}.
A recent reanalysis of the ANL and BNL data~\cite{Wilkinson14} brings both data sets closer to the original ANL one, therefore, one expects that using a fit of $C_5^A(0)$ to this new data will reduce the computed cross sections presented in this work.
%
For the $\Delta$ propagator $S_{\Delta,\alpha\beta}$, 
we use the Rarita-Schwinger prescription which depends on the $\Delta$-decay width 
%
$$\Gamma^{\text{{\tiny free}}}_{\text{{\tiny width}}}(W) = \frac{(f_{\pi N\Delta})^2}{12\pi m_\pi^2W}(p_\pi^{cm})^3(M+E_N^{cm})\,.$$
We compute the $\Delta N \pi$ decay constant $f_{\pi N\Delta}$ using 
$\Gamma^{\text{{\tiny free}}}_{\text{{\tiny width}}}(W=M_\Delta) = 120$ MeV, this results in the value $f_{\pi N\Delta}=2.21$.
Finally, for the $\Delta$ decay vertex we use $\Gamma_{\Delta\pi N}^\alpha = \frac{f_{\pi N\Delta}}{m_\pi}P_\pi^\alpha$.
%
%

Inside a nucleus, the mass and the width of the $\Delta$ resonance are modified. We use the Oset and Salcedo~\cite{Oset87} formalism to implement these medium modifications (MM):
$$\Gamma_{\text{{\tiny width}}}^{\text{{\tiny free}}} \longrightarrow \Gamma_{\text{{\tiny width}}}^{\text{{\tiny in-medium}}} = \Gamma_{\text{{\tiny Pauli}}} - 2\Im(\Sigma_\Delta) \,,\,\,\,\,\,\, M_\Delta^{\text{{\tiny free}}} \longrightarrow M_\Delta^{\text{{\tiny in-medium}}} = M^{\text{{\tiny free}}}_\Delta + \Re(\Sigma_\Delta)\,,$$
where 
%
$$-\Im(\Sigma_\Delta) = C_{QE}\left(\rho/\rho_0\right)^\alpha + C_{A2}\left(\rho/\rho_0\right)^\beta + C_{A3}\left(\rho/\rho_0\right)^\gamma\,,$$ and
$\Re(\Sigma_\Delta)=40\text{ MeV}\,(\rho/\rho_0)\,.$
%
%
Explicit expressions for $\Gamma_{\text{{\tiny Pauli}}}$ are given in Ref.~\cite{Nieves93}. Since we are working in momentum space, we do not have access to the nuclear density $\rho$. 
For this reason, we fix its value at $\rho=0.75\ \rho_0$, with $\rho_0=17$ fm$^{-3}$ the saturation density.
Finally, we modify the free $\Delta\pi N$-decay constant ($f_{\Delta\pi N}$) to take into account the $E$-dependent medium modifications:
$f_{\Delta\pi N}^{\text{\tiny in-medium}}(W) = f_{\Delta\pi N} \left[ (\Gamma_{\text{{\tiny Pauli}}} + 2C_{QE}\left(\rho/\rho_0\right)^\alpha)\ /\ \Gamma_{\text{{\tiny width}}}^{\text{{\tiny free}}} \right]^{1/2}\,.$
%

\begin{figure}[tbh]
	(a)\hspace{-0.5cm}\includegraphics[width=.217\textwidth,angle=0]{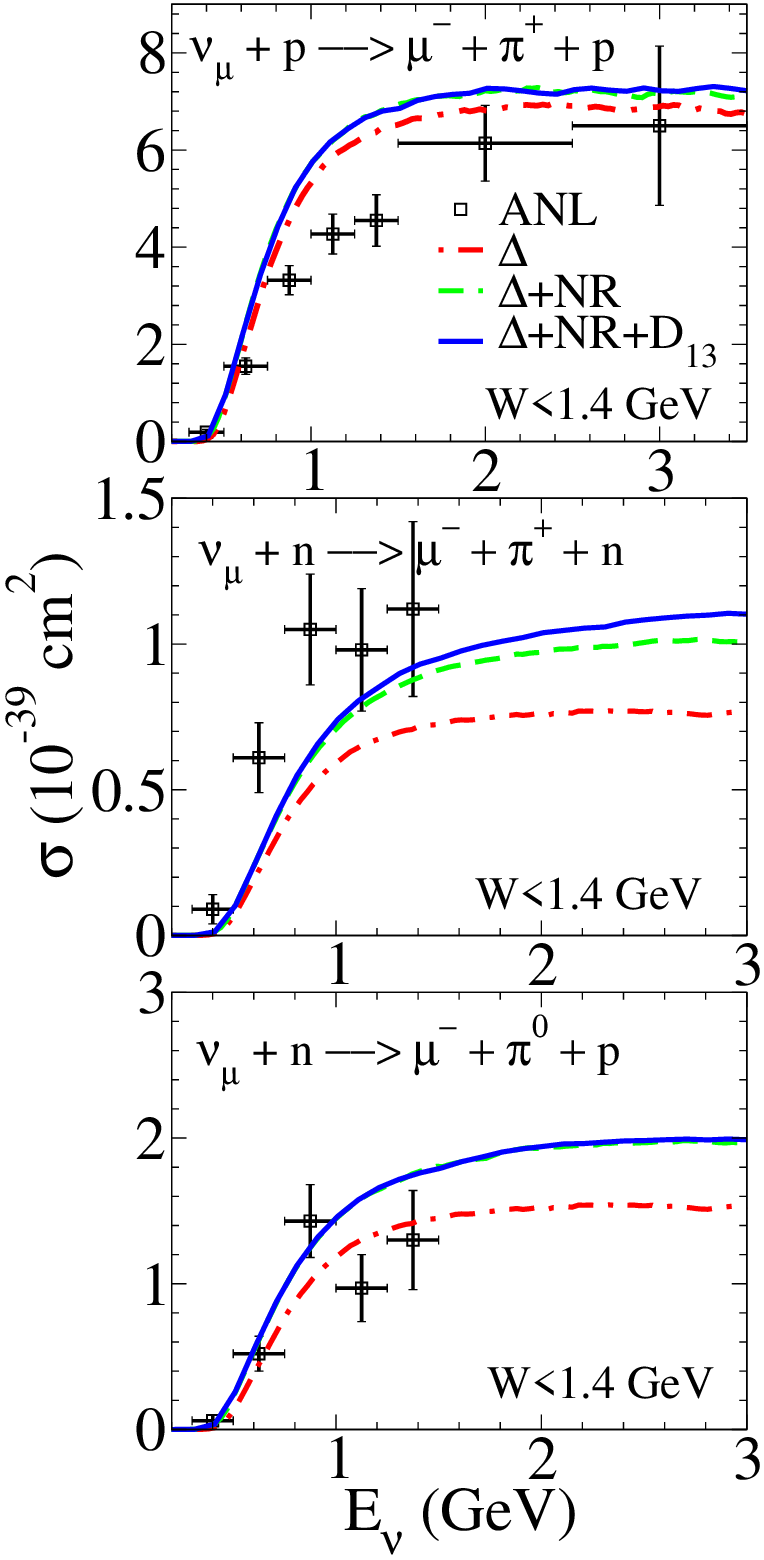}
	(b)\hspace{-0.5cm}\includegraphics[width=.2\textwidth,angle=0]{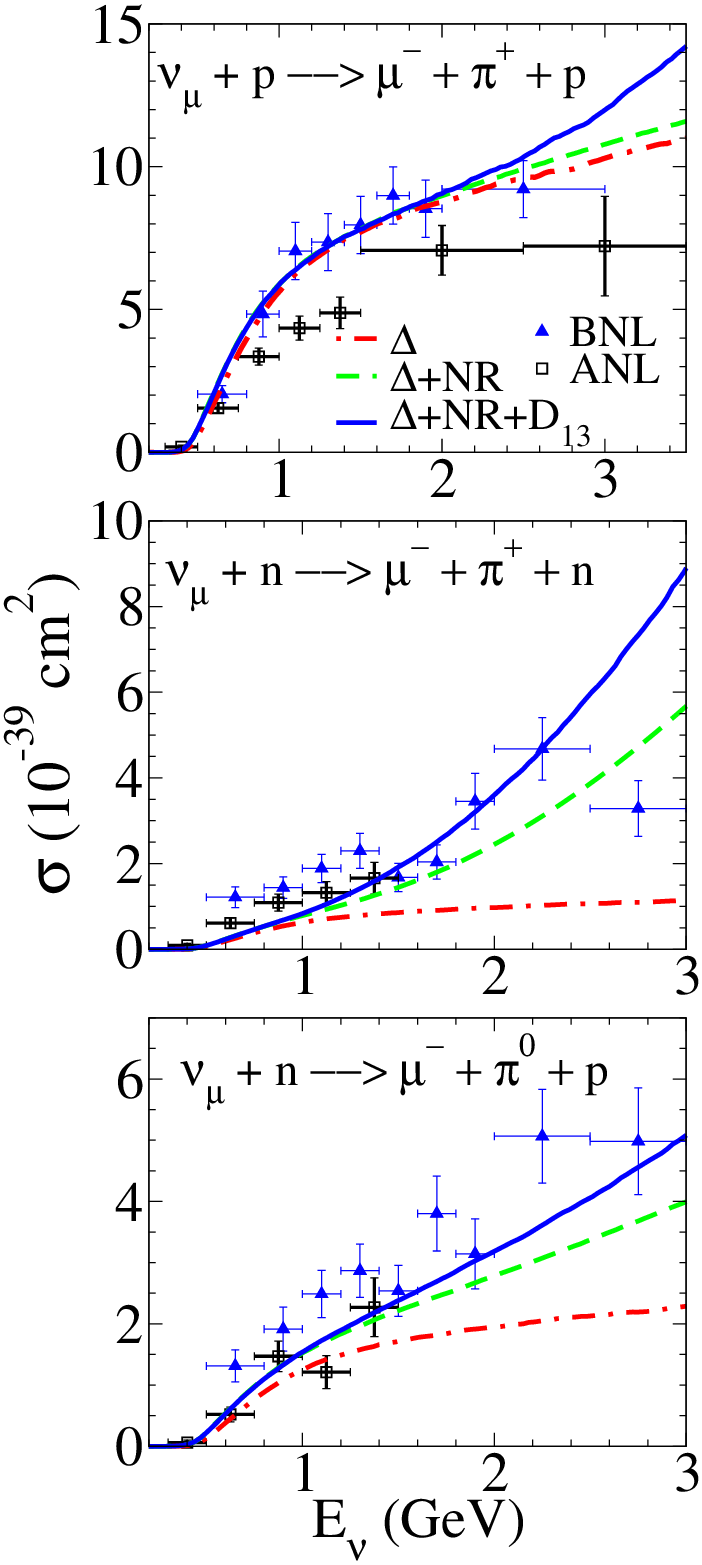}
	(c)\hspace{-0.5cm}\includegraphics[width=.6\textwidth,angle=0]{MB_total_integ.eps}
\caption{Total 1-$\pi$ production cross sections as a function of neutrino energy. In panel (a) our predictions are compared with ANL data~\cite{ANL82} where a cut in the invariant mass $W<1400$ MeV is applied in both the data and the model. 
In (b) the cut in $W$ is removed and BNL data~\cite{BNL86} are also included.
In panel (c) we confront our results with MiniBooNE data~\cite{MB11}. 
The results of Ref.~\cite{Sobczyk13} are shown as reference.}
\label{ANL-BNL}
\end{figure}

\begin{figure}[tbh]
\centering
      (a)\hspace{-0.2cm}\includegraphics[width=.31\textwidth,angle=270]{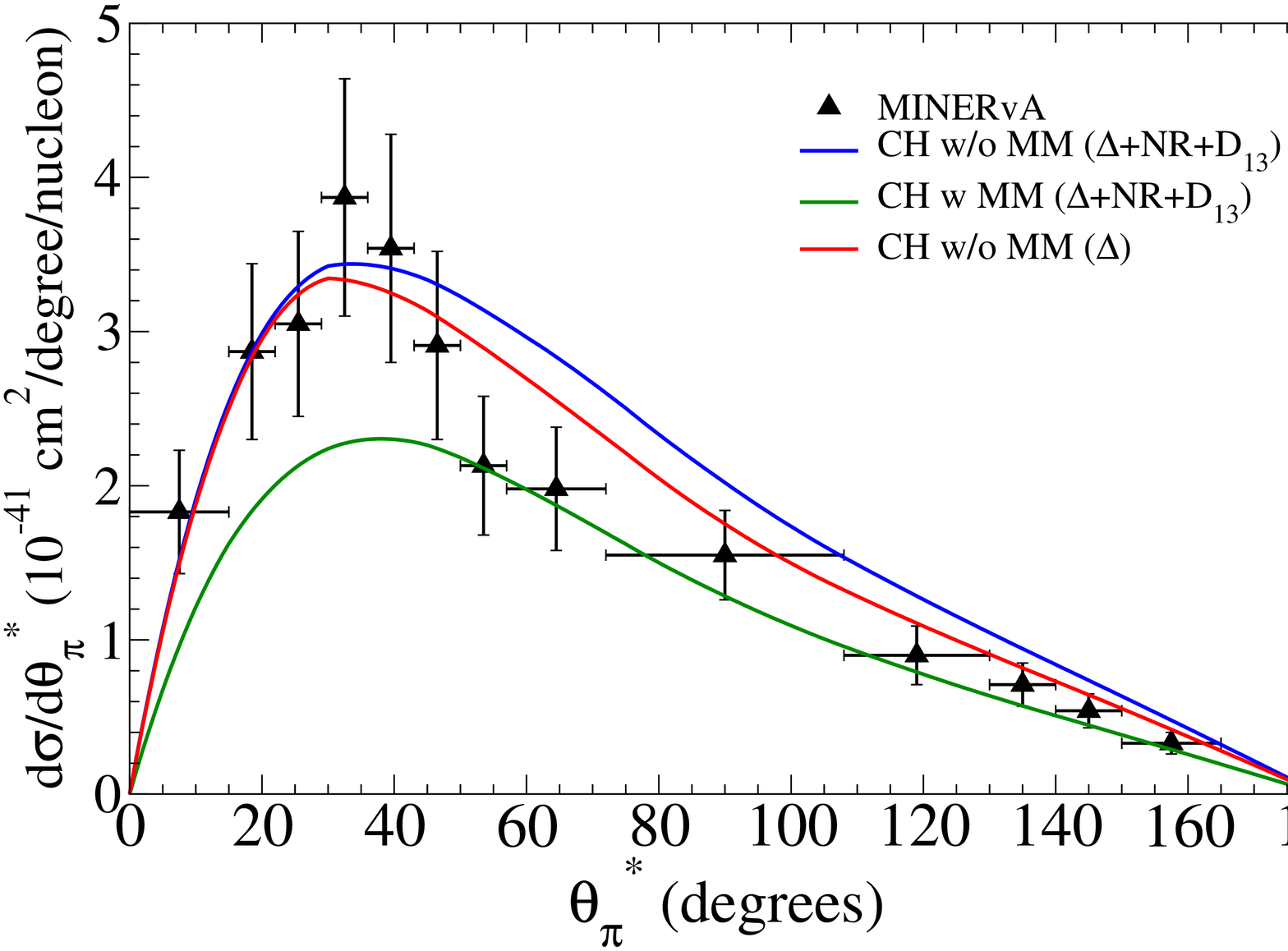}
      (b)\includegraphics[width=.31\textwidth,angle=270]{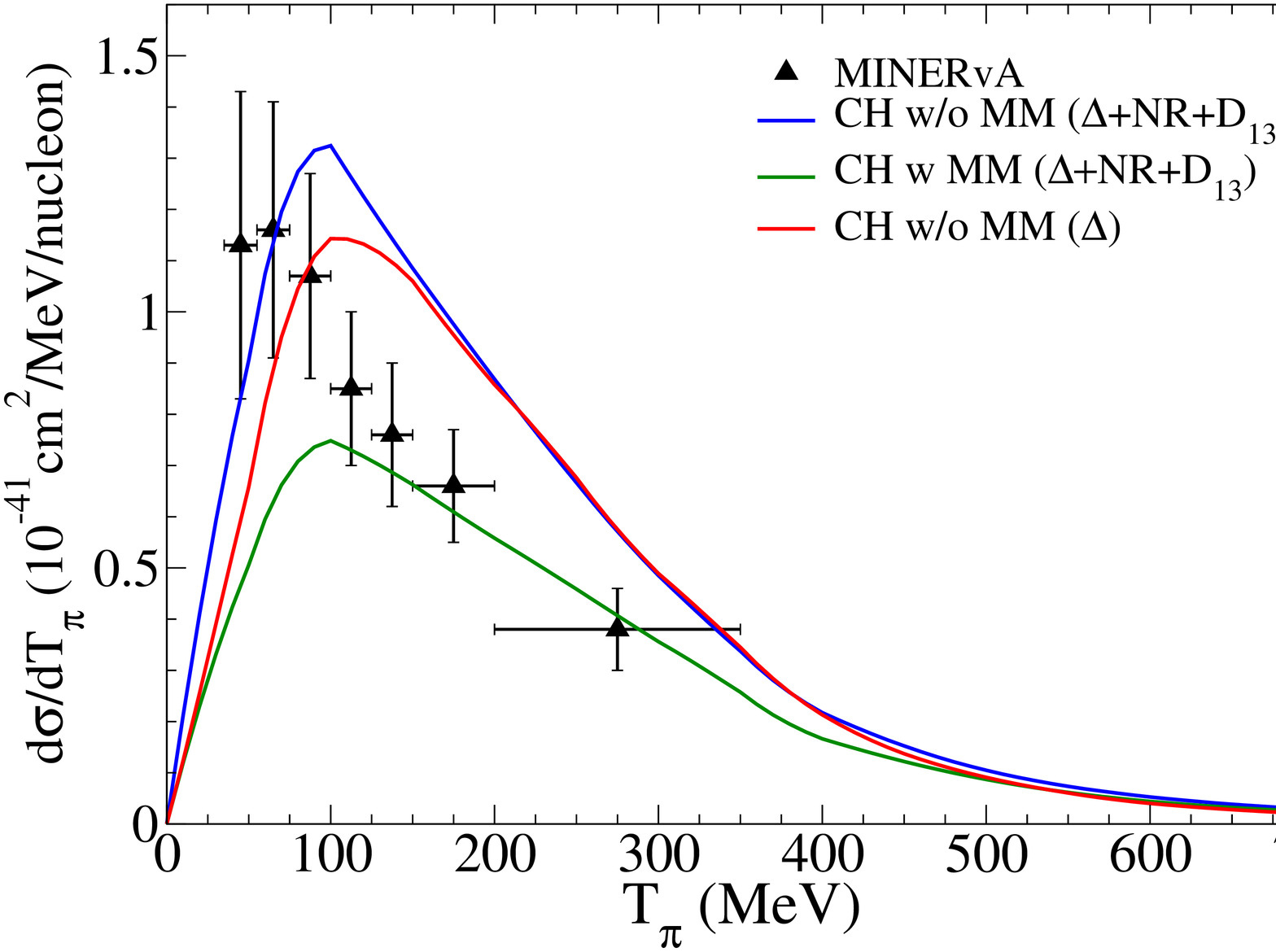}\\
%
	(c)\hspace{-0.2cm}\includegraphics[width=.31\textwidth,angle=270]{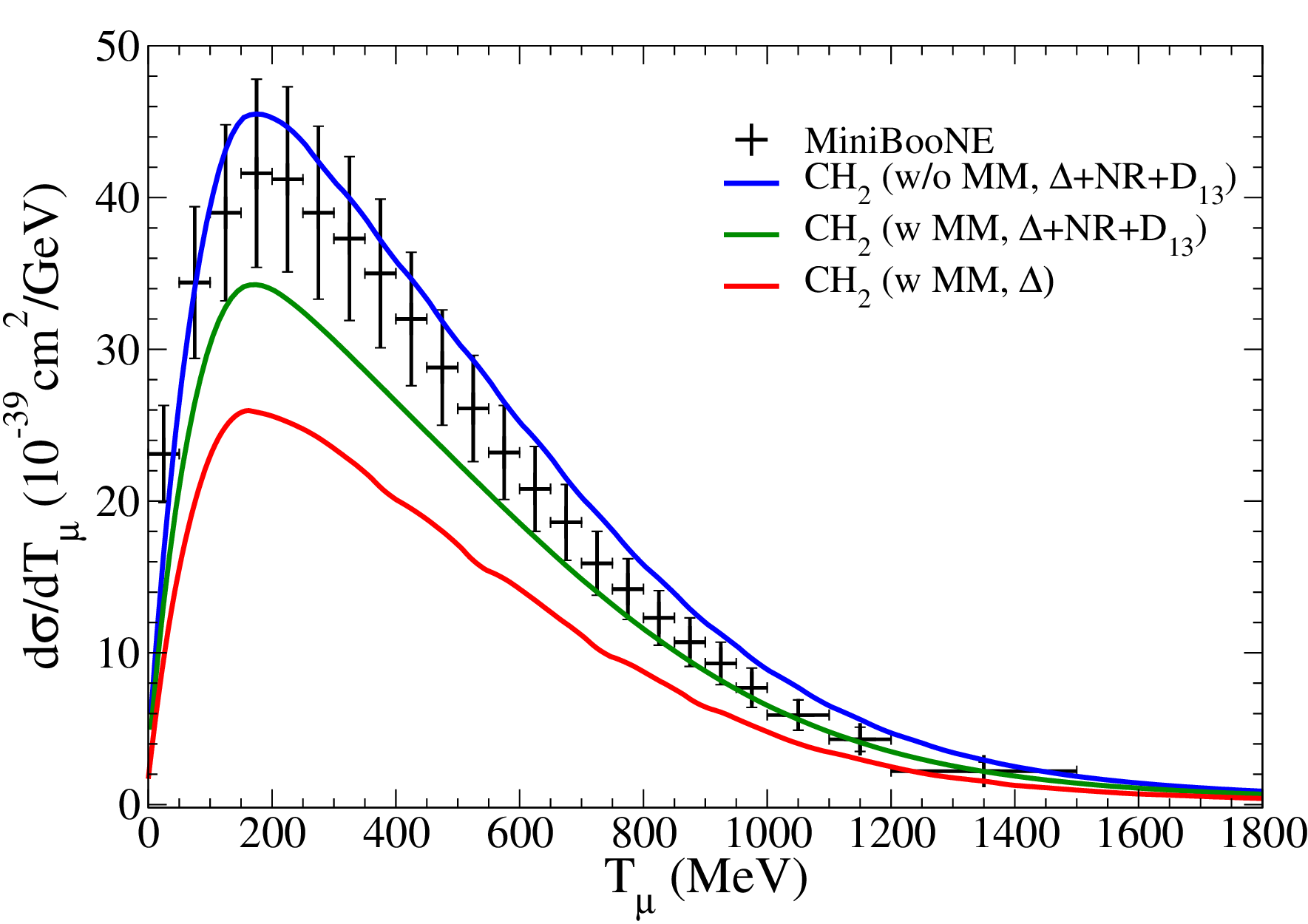}
	(d)\includegraphics[width=.31\textwidth,angle=270]{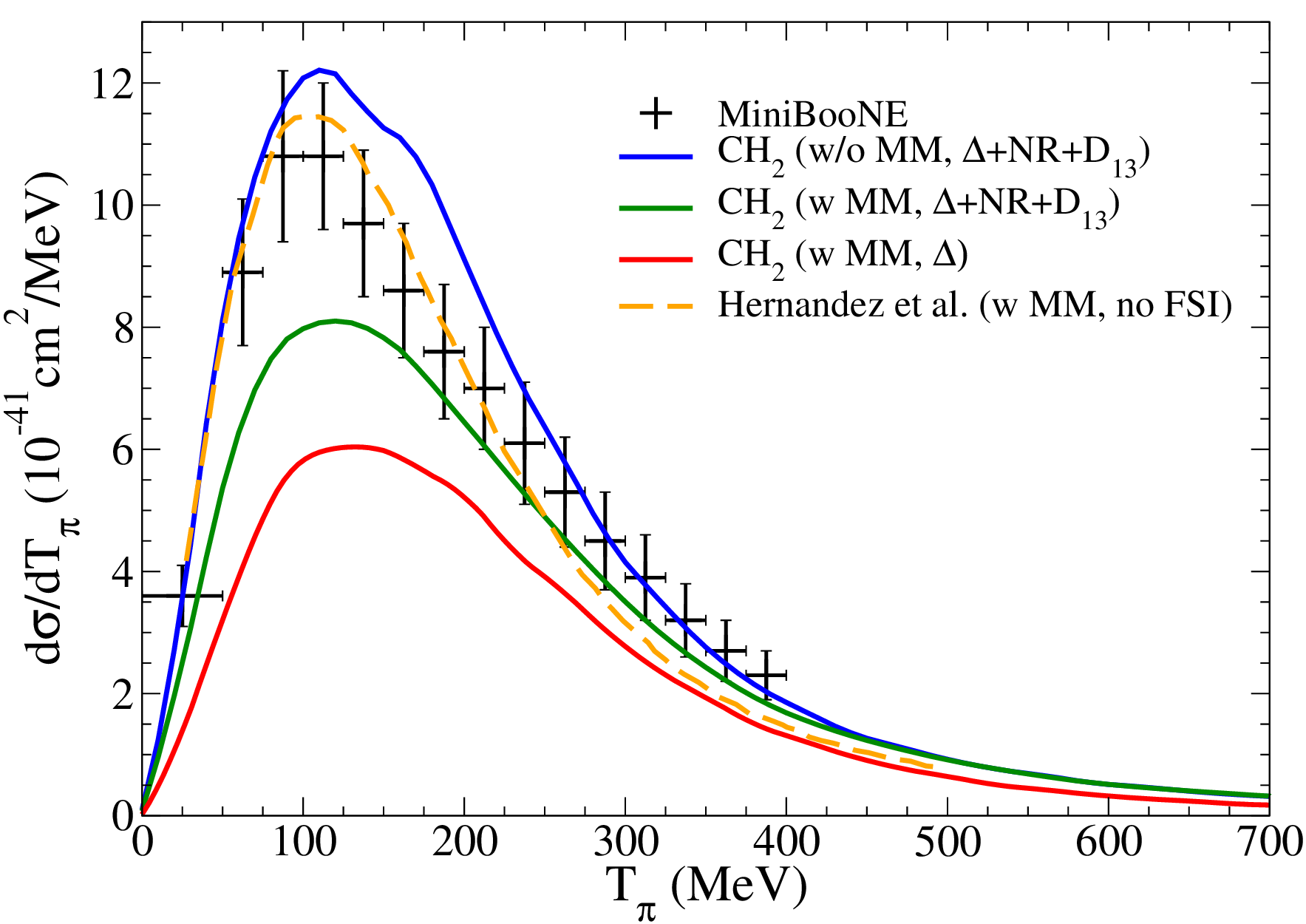}
	\caption{(a) and (b): Our predictions are compared with d$\sigma/$d$\theta_\pi^*$ and d$\sigma/$d$T_\pi$ MINER$\nu$A data~\cite{Minerva15}, respectively. (c) and (d): d$\sigma/$d$T_\mu$ and d$\sigma/$d$T_\pi$ MiniBooNE data~\cite{MB11} are compared with our predictions, respectively. In panel (d) we also show the result of Ref.~\cite{Hernandez13} as reference. }
\label{MB-Minerva}
\end{figure}

\begin{figure}[tbh]
\centering
	\hspace{-0.4cm}(a)\hspace{-0.4cm}\includegraphics[width=.44\textwidth,angle=270]{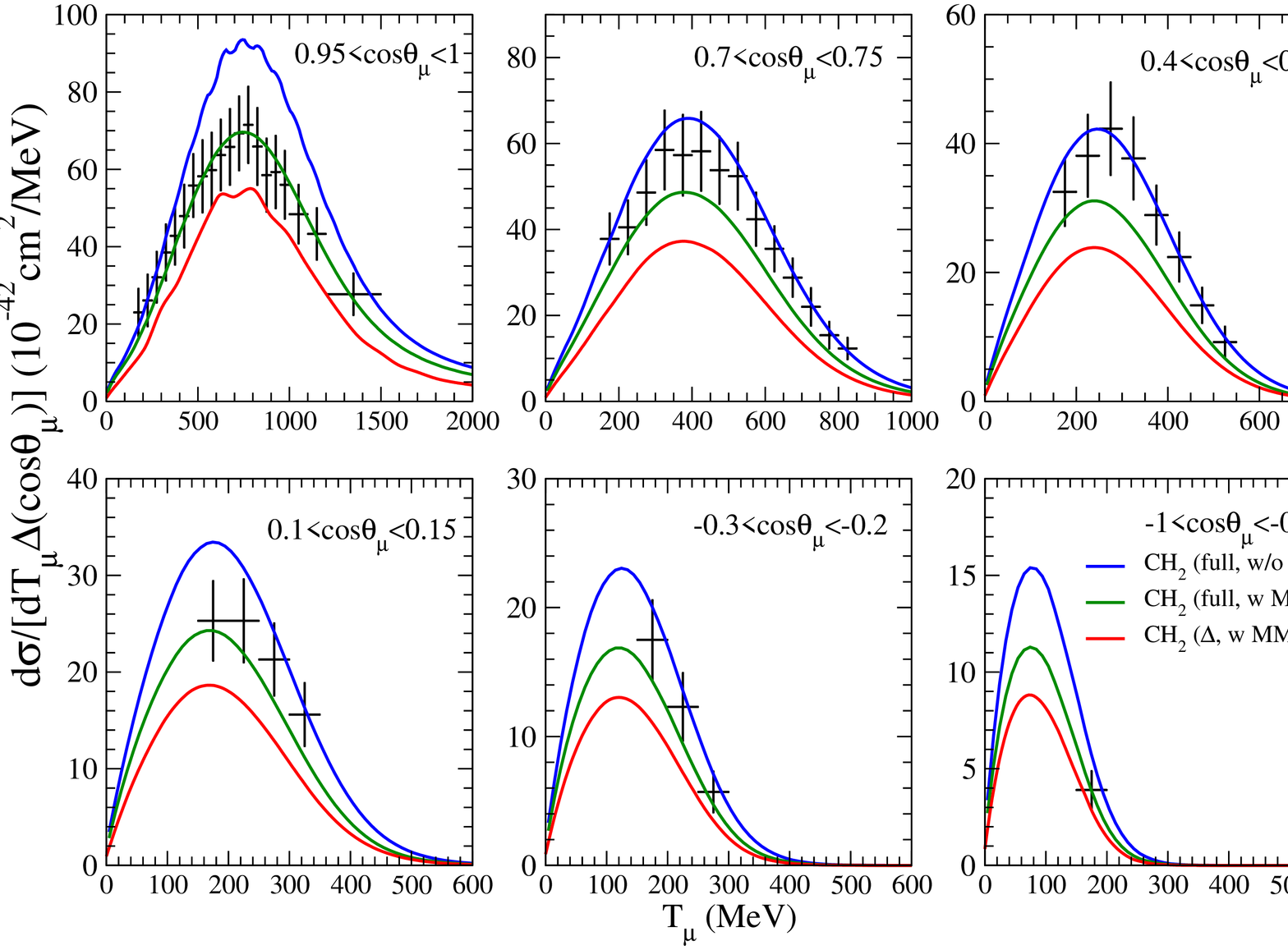}\\
	(b)\hspace{-0.5cm}\includegraphics[width=.44\textwidth,angle=270]{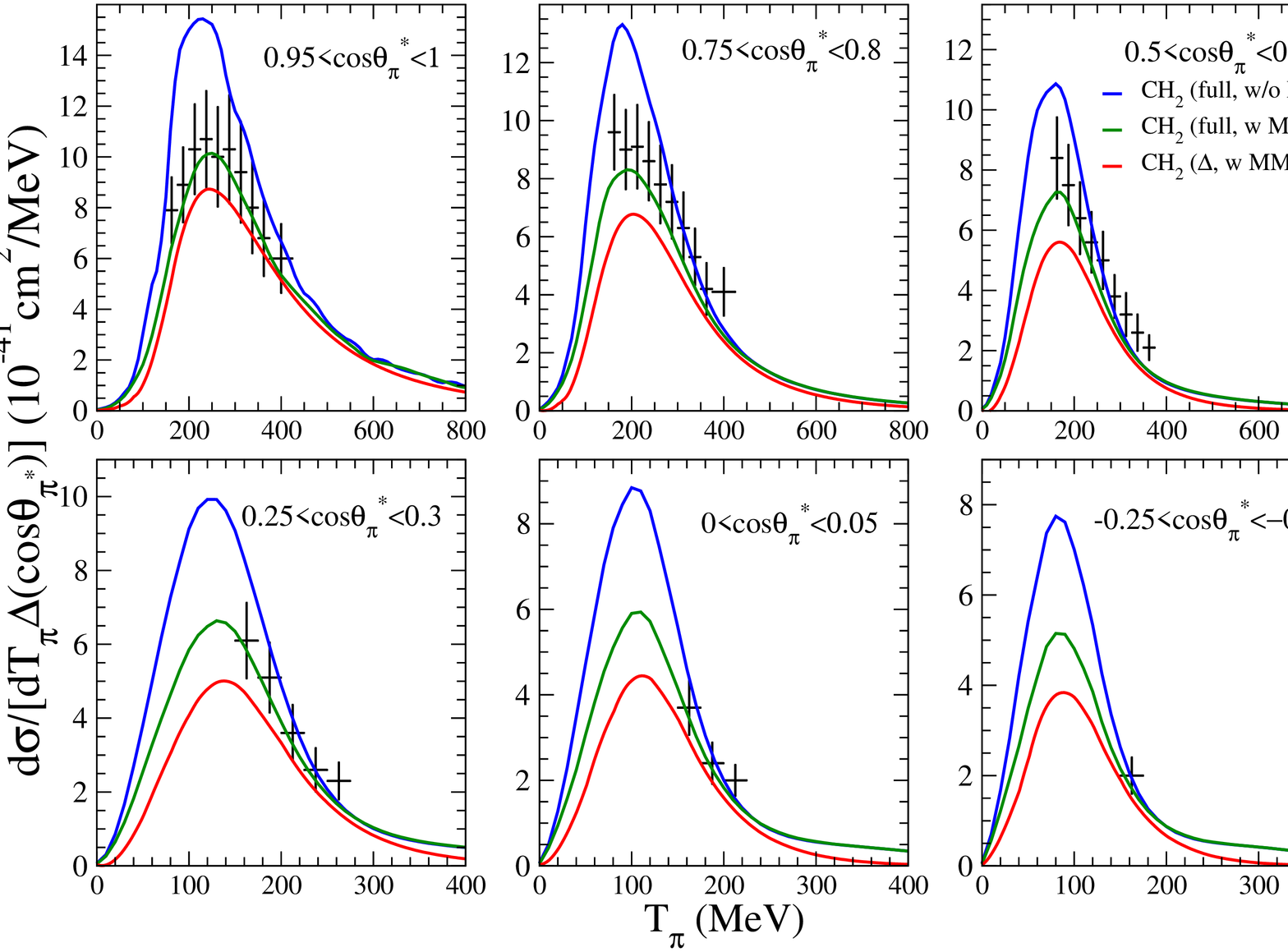}
	\caption{We compare our calculations with MiniBooNE double differential cross sections~\cite{MB11} as a function of the lepton variables (panel (a)) and pion variables (panel (b)).}
\label{MB-double}
\end{figure}

We compare our predictions with experimental data in Figs.~\ref{ANL-BNL}-\ref{MB-double}. 
%
We have studied the relative importance of various contributions to the cross sections: only $\Delta$P term (denoted by $\Delta$), $\Delta$P + C$\Delta$P + non-resonant terms (denoted by $\Delta$+NR), and the full model, including the contribution of the $D_{13}$ resonance (denoted by $\Delta$+NR+$D_{13}$ or {\it full}). 
In Fig.~\ref{ANL-BNL} (a) and (b), our model is compared with BNL and ANL neutrino-deuteron 1-$\pi$ production data. 
In this case, we neglected the medium effects and FSI, i.e., we consider scattering off free nucleons.
Our results in Fig.~\ref{ANL-BNL} (a) are consistent with those in Refs.~\cite{Hernandez07,ZmudaPhD}.
We also present the comparison of our predictions with recent MiniBooNE and MINER$\nu$A data. 
The total MiniBooNE cross section is presented in Fig.~\ref{ANL-BNL} (c). 
Single and double differential MiniBooNE and MINER$\nu$A cross sections for a variety of kinematics are shown in Figs.~\ref{MB-Minerva} and \ref{MB-double}. 
In this case, we show $\Delta$P and {\it full} model cross sections. Also, the effect of MM on the cross sections is investigated.
%
%

\vspace{0.4cm}
Our main conclusions can be summarized as follows:
   \begin{itemize}
    \item  In general, the shape of the data is well reproduced by the model.
    \item  Non-resonant and $D_{13}$ contributions are essential.
    \item  MM reduce the $\Delta$ cross sections by approximately 40-50\%. This results in a reduction of the full-model cross sections of approximately 20\%.
    \item  When MM are considered, our results are notably lower than Hernandez et al.~\cite{Hernandez13} (Fig.~\ref{MB-Minerva} (d)) and Sobczyk and \.Zmuda~\cite{Sobczyk13} (Fig.~\ref{ANL-BNL} (c)). Also, we underpredict both MiniBooNE and MINER$\nu$A data. 
    This suggests that we should explore different ways of implementing MM.
    Work on this is in progress.
    \item  Other contributions to the cross sections such as the excitation of other resonances, coherent pion production and processes involving more than 1$\pi$ in the final state may improve the agreement with data.
   \end{itemize}



\begin{thebibliography}{9}
%
\bibitem{MB11}
A. A. Aguilar-Arevalo et al. (MiniBooNE Collaboration), Phys Rev. D {\bf 83}, (2011) 013005.
%
\bibitem{Minerva15}
B. Eberly et al. (MINER$\nu$A Collaboration), Phys. Rev. D. {\bf 92}, (2015) 092008.
%
\bibitem{Hernandez07}
E. Hern\'andez, J. Nieves and M. Valverde, Phys. Rev. C {\bf 76}, (2007) 033005.
%
\bibitem{Hernandez13}
E. Hern\'andez, J. Nieves and M. J. {Vicente Vacas}, Phys. Rev. D {\bf 87}, (2013) 113009.
%
\bibitem{Praet09}
C. Praet, O. Lalakulich, N. Jachowicz, and J. Ryckebusch, Phys. Rev. C {\bf 79}, (2009) 044603.
%
\bibitem{Sobczyk13}
Jan T.~Sobczyk and Jakub \.Zmuda, Phys. Rev. C {\bf 87}, (2013) 065503.
%
\bibitem{Leitner09}
T. Leitner, O. Buss, L. Alvarez-Ruso, and U. Mosel, Phys. Rev. C {\bf 79}, (2009) 034601.
%
\bibitem{Lalakulich06} 
Olga Lalakulich, Emmanuel A. Paschos, and Giorgi Piranishvili, Phys. Rev. D {\bf 74}, (2006) 014009.
%
\bibitem{BNL86}
T. Kitagaki et al., Phys. Rev. D {\bf 34}, (1986) 2554.
%
\bibitem{Paschos04}
 E. A. Paschos, J.-Y. Yu, and M. Sakuda, Phys. Rev. D {\bf 69}, (2004) 014013.
%
\bibitem{Wilkinson14}
Callum Wilkinson et al., Phys. Rev. D {\bf 90}, (2014) 112017.
%
\bibitem{Oset87}
E. Oset and L. L. Salcedo, Nucl. Phys. A {\bf 468}, (1987) 631.
%
\bibitem{Nieves93}
J. Nieves, E. Oset, and C. Garcia-Recio, Nucl. Phys. A {\bf 554}, (1993) 554.
%
\bibitem{ANL82}
G. M. Radecky et al., Phys Rev. D {\bf 25}, (1982) 1161.
%
\bibitem{ZmudaPhD}
Jakub \.Zmuda, PhD Thesis, University of Wroclaw (2014)
%
\end{thebibliography}
\end{document}